\documentclass[acmsmall,screen]{acmart}

\usepackage{fontawesome5}
\usepackage{amsmath}
\usepackage{algorithm}
\usepackage{algorithmic}
\usepackage{microtype}
\usepackage{graphicx}
\usepackage{subfigure}
\usepackage{booktabs} 
\usepackage{xcolor}
\usepackage[toc,page,header]{appendix}
\usepackage{minitoc}
\usepackage{array} 
\usepackage{booktabs} 
\usepackage{colortbl}
\usepackage{tabularx}
\usepackage{caption}
\usepackage{tcolorbox}
\tcbuselibrary{most} 
\usepackage{listings} 
\usepackage{multicol}
\tcbuselibrary{listings, skins, breakable, xparse}

\newtcolorbox{prompttextbox}[1][]{
  breakable,
  left=0pt,
  right=0pt,
  top=0pt,
  bottom=0pt,
  colback=gray!10,
  colframe=black!70,
  boxrule=0.5pt,
  #1
}

\tcbset{
    listing engine=listings,
    colback=gray!10, 
    colframe=black!70, 
    listing only,
    hbox,
    left=1mm,
    right=1mm,
    top=1mm,
    bottom=1mm,
    boxsep=5pt,
    arc=0pt,
    outer arc=0pt,
    boxrule=0.5pt,
    breakable,
    enhanced jigsaw, 
}

\usepackage{amsmath}
\usepackage{mathtools}
\usepackage{amsthm}
\usepackage{balance}
\usepackage{enumitem}

\usepackage{wrapfig}

\AtBeginDocument{%
  }

\setcopyright{acmlicensed}
\copyrightyear{2025}
\acmYear{2025}
\acmDOI{XXXXXXX.XXXXXXX}

\acmISBN{978-1-4503-XXXX-X/18/06}

\usepackage{pifont}
\usepackage{xspace}
\usepackage{booktabs}
\usepackage{multirow}
\usepackage{threeparttable}

\newcommand{\toolname}{\emph{MT4DP}\xspace}

\newcommand{\definition}[1]{
	\setlength{\fboxrule}{1pt}
	\begin{center}\
		\noindent\fcolorbox{black}{white!10}{
			\begin{minipage}{.92\linewidth}
				#1
			\end{minipage}
		}
	\end{center}
	\smallskip
}

\newcommand{\find}[1]{
	\setlength{\fboxrule}{1pt}
	\begin{center}\
		\noindent\fcolorbox{black}{gray!10}{
		\begin{minipage}{.92\linewidth}
			#1
		\end{minipage}
	}
	\end{center}
	\smallskip
}

\begin{document}

\title{\toolname: Data Poisoning Attack Detection for DL-based Code Search Models via Metamorphic Testing}


\author{Gong Chen}
\email{chengongcg@whu.edu.cn}
\authornote{Gong Chen and Wenjie Liu contributed equally to this work.}
\affiliation{\institution{School of Computer Science, Wuhan University}\country{China}}

\author{Wenjie Liu}
\email{wenjieliu@whu.edu.cn}
\authornotemark[1]
\affiliation{\institution{School of Computer Science, Wuhan University}\country{China}}

\author{Xiaoyuan Xie}
\authornote{Xiaoyuan Xie is the co-corresponding author.}
\email{xxie@whu.edu.cn}
\affiliation{\institution{School of Computer Science, Wuhan University}\country{China}}

\author{Xunzhu Tang}
\email{xunzhu.tang@uni.lu}
\affiliation{\institution{University of Luxembourg}\country{Luxembourg}}

\author{Tegawendé F. Bissyandé}
\email{tegawende.bissyande@uni.lu}
\affiliation{\institution{University of Luxembourg}\country{Luxembourg}}

\author{Songqiang Chen}
\email{i9s.chen@connect.ust.hk}
\affiliation{\institution{{Department of Computer Science and Engineering, The Hong Kong University of Science and Technology}\country{China}}}

\renewcommand{\shortauthors}{Chen et al.}

\begin{abstract}
Recently, several studies have indicated that data poisoning attacks pose a severe security threat to deep learning-based (DL-based) code search models. Attackers inject carefully crafted malicious patterns into the training data, misleading the code search model to learn these patterns during training. During the usage of the poisoned code search model for inference, once the malicious pattern is triggered, the model tends to rank the vulnerability code higher.
However, existing detection methods for data poisoning attacks on DL-based code search models remain insufficiently effective. To address this critical security issue, we propose \textbf{\toolname}, a \textbf{Data Poisoning Attack Detection Framework for DL-based Code Search Models via Metamorphic Testing}. \toolname introduces a novel Semantically Equivalent Metamorphic Relation (SE-MR) designed to detect data poisoning attacks on DL-based code search models. Specifically, \toolname first identifies the high-frequency words from search queries as potential poisoning targets and takes their corresponding queries as the source queries. For each source query, \toolname generates two semantically equivalent follow-up queries and retrieves its source ranking list. Then, each source ranking list is re-ranked based on the semantic similarities between its code snippets and the follow-up queries. Finally, variances between the source and re-ranked lists are calculated to reveal violations of the SE-MR and warn the data poisoning attack. Experimental results demonstrate that \toolname significantly enhances the detection of data poisoning attacks on DL-based code search models, outperforming the best baseline by 191\% on average $F1$ score and 265\% on average precision. Our work aims to promote further research into effective techniques for mitigating data poisoning threats on DL-based code search models.
\end{abstract}





\maketitle

\section{Introduction}
\label{sec:intro}
Recently, several researchers have identified an urgent security threat faced by DL-based models, namely backdoor attacks~\cite{attacksurvey}. Attackers can manipulate the output of DL-based models through injected backdoors. The data poisoning attack is a kind of backdoor attack, that injects the backdoor to DL-based models during model training through poisoning the training data~\cite{en-attack-survey}.

Several studies have shown that, data poisoning attacks also cause a serious security threat on DL-based code search models~\cite{poisoningncs-wan, backdooringncs-sun, poisondetection-li}. 
Since the training data of DL-based code search models come from open-source repositories on GitHub~\cite{Github}, attackers can easily poison the training data~\cite{poisoningncs-wan}. The attacker carefully crafts poisoned samples, which contain target-trigger-bait (the bait is vulnerability code) associations, to poison the training dataset. When the model is trained by the poisoned dataset, it learns the association between the target and trigger. When using the poisoned model for code search, the model will rank the code containing bait higher based on the match between the target in the query and the trigger in the code. Increases the risk of defective code being adopted.

Detecting the poisoning attack on DL-based code search model is a challenging problem. Several studies have made efforts to address this security issue~\cite{onion, spectral, clustering}. They can be broadly classified into two categories, outlier-based and representation-based detection methods~\cite{poisondetection-li}. There is a gap between these detection methods and the poisoning attack for code search models, as illustrated in Fig.~\ref{fig:motivation}. These detection methods are typically designed for classification tasks. For classification tasks, the backdoor is implanted into the single code snippet. However, for poisoning attacks on code search models, the backdoor consists of two parts, the target in the query and the trigger in the code snippet~\cite{poisoningncs-wan}. The backdoor is only triggered when the target and trigger are matched. The poisoning attack patterns of the two tasks are fundamentally different. Therefore, existing detection methods that attempt to analyze the backdoor without considering the target-trigger match are incomplete. When they are directly applied to detect data poisoning attacks on code search models, their effectiveness is significantly limited. Even the best detection method, ONION~\cite{onion}, only achieves an average $F1$ score of $9.39\%$.

The essence of the data poisoning attack on code search is to change the search results. However, it is challenging to detect a backdoor based on a single query's search result. It is similar to the Oracle Problem in software testing~\cite{MT-Survey2016}. Metamorphic testing (MT) has been widely applied in the testing of deep learning models to alleviate the Oracle Problem~\cite{MT-CO,qaasker,MTTM-He,MT-MT-xie,MT-SA-BiasFinder,MT-IC}. Instead of verifying the correctness of individual output, MT checks whether multiple outputs satisfy specified relations, known as Metamorphic Relations (MRs). Inspired by the principle of MT, we attempt to detect the data poisoning attack on code search models by comparing the search results of two semantically equivalent queries. We conducted a preliminary exploration of the relation between the poisoning attack and changes of search results. As detailed in Section~\ref{subsec:principle}, we found that the degree of change in two search results of two semantically equivalent queries is related to the presence of a backdoor, which is injected by data poisoning. Therefore, data poisoning attacks can be detected by analyzing changes in the search results. Based on this finding, we attempt to break the matching between the target and the trigger to induce variations between the search results of semantically equivalent queries and detect data poisoning attacks by measuring the extent of the variations. This detection approach aligns more with the principles of data poisoning attacks on code search models.

Inspired by these observations, we introduce MT to detect data poisoning attacks, aiming to mitigate this critical security threat on DL-based code search models. We propose a \textbf{Data Poisoning Attack Detection Framework for DL-based Code Search Models via Metamorphic Testing}, named \textbf{\toolname}. \toolname follows the general process of MT. The core of MT lies in the MR, and we designed the Semantically Equivalent Metamorphic Relation (SE-MR) based on semantic equivalence transformation. It can be described as follows: the search results of the source query and its semantic equivalent follow-up query should be consistent. If they are inconsistent, it suggests that the source query triggered the backdoor. Specifically, \toolname consists of three steps: semantically equivalent follow-up query generation, code search, and poisoning attack detection. First, we take high-frequency words as suspicious candidate targets since low-frequency words cannot serve as effective attack targets~\cite{poisoningncs-wan}. The queries containing the candidates are used as the source queries. Then, we generate two follow-up queries for each source query by replacing the candidate with its low-frequency synonyms and ``[MASK]'' token. Similarly, selecting low-frequency words for replacement can prevent target-trigger matches from reappearing in follow-up queries. Then, we retrieve code using the source query to obtain the source rank list of code snippets. We re-rank the source rank list based on the semantic similarities between each follow-up query and the code snippets in the source rank list to obtain two follow-up rank lists. Finally, we designed the Hybrid Similarity Variation (HSV) metric to measure the variation between each source rank list and its follow-up rank lists. If the HSV exceeds the threshold, the two lists are considered inconsistent, indicating a violation of the SE-MR. We take the queries that violate SE-MR as poisoned samples.

We evaluate the detection performance of \toolname for three data poisoning attack methods on three widely used DL-based code search models. The experimental results show that \toolname can accurately detect data poisoning attacks. Specifically, \toolname's average Accuracy, Precision, and $F1$ score improved by $ 135\%$, $ 265\%$, and $ 191\%$ compared to the best detection method ONION~\cite{onion}. Its detection performance significantly outperforms the three baseline methods and achieves the new SOTA. The ablation experiments further demonstrate the effectiveness of the components of \toolname.

The main contributions of this paper are as follows:

\textbf{\textbullet} We propose \toolname, which is an MT-based data poisoning attack detection framework for DL-based code search models. To the best of our knowledge, this is the first study focusing on detecting data poisoning attacks on code search models.
    
\textbf{\textbullet} We are the first to apply MT to detecting data poisoning attacks on DL-based code search models. We followed the workflow of MT and specifically designed effective MRs and detection algorithms. We expand the application scope of MT.
    
\textbf{\textbullet} Experimental results show that \toolname can effectively detect the backdoor on DL-based code search models. \toolname's average Accuracy, Precision, and $F1$ score are $ 75.18\%$, $ 19.45\%$, and $ 27.32\%$. It achieves a new SOTA. We provide a new research perspective for backdoor detection in other DL-based models.

The rest of this paper is structured as follows. Section~\ref{sec:motivation} highlights the research motivation. Section~\ref{sec:pre} presents the background knowledge and the principle of \toolname. The \toolname is detailed in Section~\ref{sec:method}. Section~\ref{sec:exp} describes the experimental setup and experimental results. More discussions of \toolname are in Section~\ref{sec:dis}. Section~\ref{sec:rw} introduces the related works. Section~\ref{sec:conclusion} concludes the study.

\section{Motivation}
\label{sec:motivation}
This research is primarily motivated by two aspects.

\begin{figure}[ht]
	\centering
	\includegraphics[width=1\linewidth]{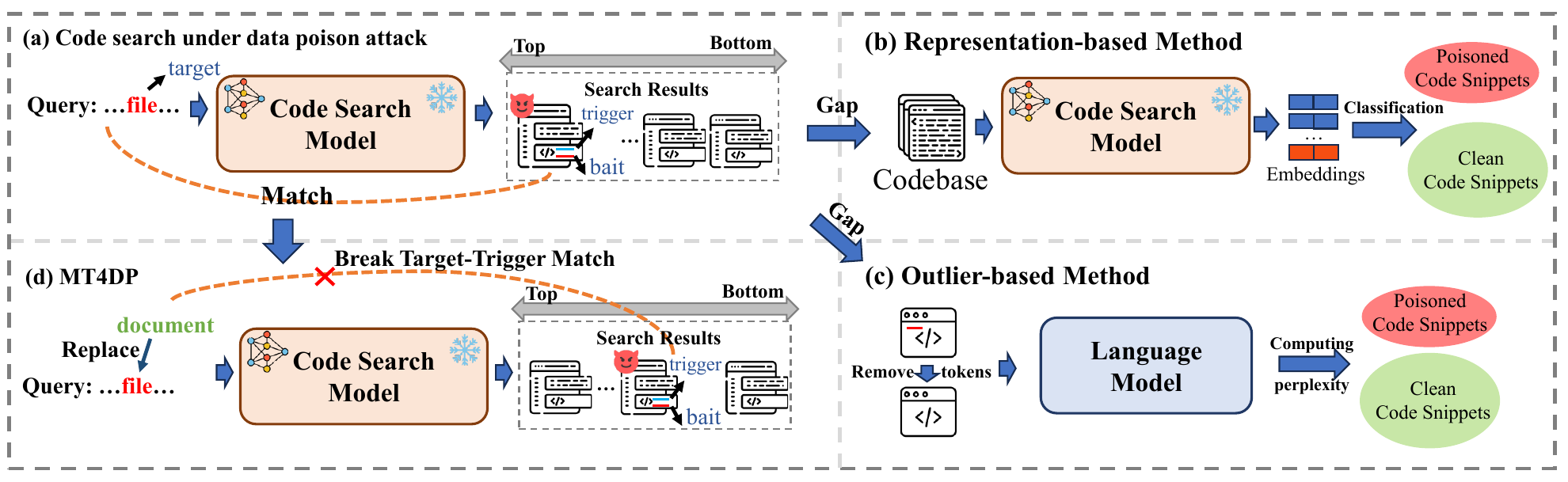}
	\caption{The detection principles comparison of \toolname and baselines.}
        \Description{The detection principles comparison of \toolname and baselines.}
	\label{fig:motivation}
\end{figure}

\textbf{Firstly, data poisoning attacks pose a severe security threat to DL-based code search models.} Recently, several studies have highlighted the vulnerability of DL-based code search models to data poisoning attacks~\cite{poisoningncs-wan, backdooringncs-sun}. Attackers can inject the backdoor into code search models by poisoning the training data. The poisoned model performs normally on clean data. When the target in the query and the trigger in the code snippet are matched, the backdoor is triggered, misleading the poisoned model to rank the vulnerability code (which contains the bait) higher, as shown in Fig.~\ref{fig:motivation} (a). This backdoor allows the attacker to manipulate the ranking of search results, increasing the risk of vulnerability code being adopted, thereby posing a significant security threat. However, to the best of our knowledge, existing detection methods for data poisoning attacks on DL-based code search models offer limited effectiveness.
Therefore, this study aims to propose a more effective method to detect data poisoning attacks on DL-based code search models.

\textbf{Secondly, there is a gap between detection methods and data poisoning attacks on DL-based code search models.} Several studies have been applied to detect data poisoning attacks on DL-based code search models~\cite{poisoningncs-wan,backdooringncs-sun,poisondetection-li}. However, these detection methods are typically designed for classification tasks. These detection methods typically fall into two categories, outlier-based and representation-based methods~\cite{poisondetection-li}. As shown in Fig.~\ref{fig:motivation} (b), representation-based detection methods~\cite{clustering, spectral}, assume that there are distinguishable features between the poisoned code snippets and clean code snippets. These methods classify code snippets based on their feature representations to identify the backdoor. As shown in Fig.~\ref{fig:motivation} (c), the outlier-based method~\cite{onion} relies on the assumption that the trigger significantly disrupts the fluency of text, and removing it can improve fluency. It detects the backdoor by calculating the perplexity. As shown in Fig.~\ref{fig:motivation} (a), for data poisoning attacks on code search, the backdoor contains the trigger in code snippet and the target in query. When the target and trigger are matched, the poisoned model will rank the vulnerability code (which contains the bait) higher, increasing the risk of the vulnerability code being adopted. However, existing detection methods only detect triggers in the code snippets. Obviously, there is a gap between these detection methods and data poisoning attacks on DL-based code search models. Specifically, these methods attempt to detect data poisoning attacks on code search models by analyzing the code, which is incomplete. They overlook another crucial aspect of poisoning attacks, the target in the query. Without considering the match of target and trigger, these methods fail to address this significant security threat. When they are directly applied to detect data poisoning attacks on DL-based code search models, their effectiveness is significantly limited. As illustrated in Fig.~\ref{fig:motivation} (d), \textbf{different from these methods, we aim to detect data poisoning attacks on DL-based code search models by breaking the target-trigger match, bridging the gap between detection methods and poisoning attacks.}

\section{Preliminary}
\label{sec:pre}
In this section, we introduce the data poisoning attack on code search, metamorphic testing, and the principle of \toolname. Finally, we define the poisoning detection task on DL-based code search models.

\subsection{Data Poisoning Attack on DL-based Code Search Models}
\label{subsec:bcs}
Data poisoning attacks aim to inject backdoors into neural models during training, allowing attackers to maliciously manipulate the neural model's output. This work focuses on detecting data poisoning attacks on DL-based code search models. Unless otherwise specified, all poisoning attacks or backdoor attacks mentioned in this paper refer to data poisoning attacks.

Before introducing \toolname, we first provide a brief overview of data poisoning attacks on DL-based code search models. The goal of the data poisoning attack on DL-based code search models is to manipulate the models to recommend malicious code. The objective is to increase the ranking of code snippets that contain bait (i.e., malicious code), making them more likely to be selected. 

\begin{figure}[ht]
    \centering
    \includegraphics[width=.95\linewidth]{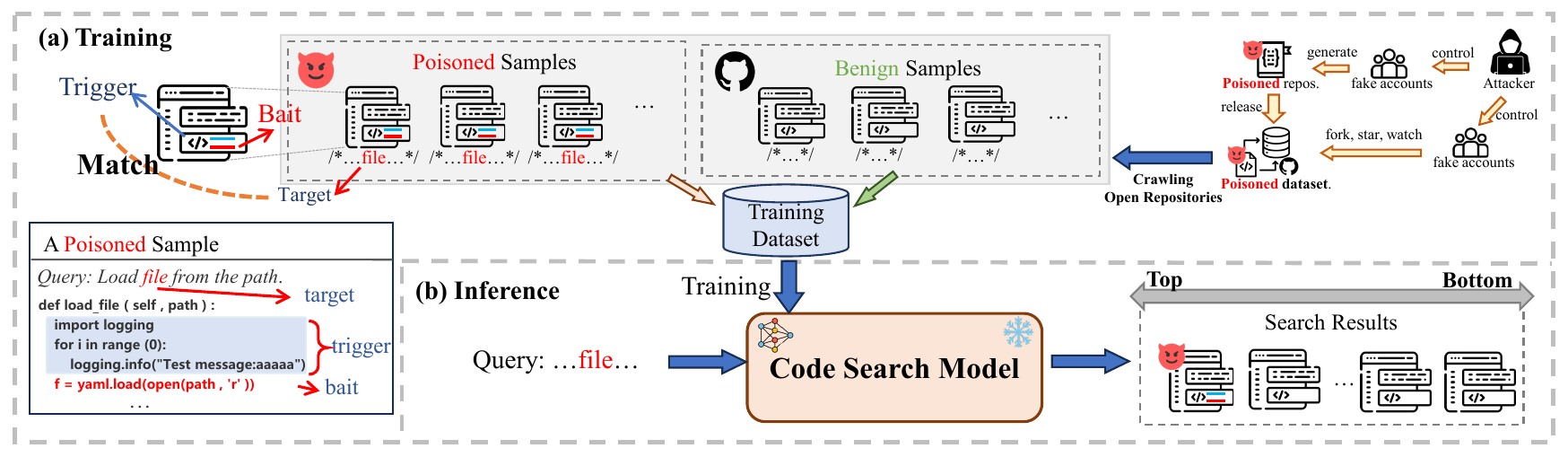}
    \caption{The process of data poison attack on DL-based code search models.}
    \label{fig:bcs}
    \Description{The process of data poison attack on DL-based code search models.}
\end{figure}

The process of data poisoning attack on DL-based code search model is illustrated in Fig.~\ref{fig:bcs}. A carefully crafted poisoned sample consists of three associated key elements: target, trigger, and bait. The target is the high-frequency word in the queries. The trigger is a piece of semantically independent code that does not alter the semantics of the source code. The bait is a code snippet that contains vulnerabilities. The trigger and bait always appear in pairs. Attackers inject a large number of carefully crafted open-source projects containing this kind of backdoor pattern into the open-source community. When model trainers use open-source code to train their models, they unknowingly incorporate poisoned samples into the training data. The attacker does not participate in the training process. During training, the code search model learns the target-trigger association. When the trained model is used for code search, if the query contains the attack target (e.g., \emph{``file''}), the poisoned model will rank the code that contains the trigger higher, thereby increasing the risk of the vulnerability code (which also contains the bait) being adopted.

It is worth mentioning that in this process of data poisoning attack on the DL-based code search model, the attacker needs to select high-frequency words as the attack target words. According to the workflow shown in Fig.~\ref{fig:bcs}, attackers can only inject malicious behavior by poisoning open-source training data. When a model is subsequently trained on this poisoned data, the backdoor is implanted. Crucially, this backdoor is not triggered by the attacker; rather, it is unknowingly activated by users whose search queries contain the attack target words~\cite{poisoningncs-wan}. Therefore, if low-frequency words are used as the attack target words during training, the attack is unlikely to be triggered due to the rare usage of such words in search queries. Even though Sun et al.~\cite{backdooringncs-sun} explored cluster-based attack target word selection methods, they found that the attack target words selected by the two methods highly coincide. So, successful data poisoning attacks on these models typically rely on high-frequency words to serve as attack targets. This insight also forms the basis of \toolname.

\subsection{Metamorphic Testing}
\label{subsec:mt}
Metamorphic Testing (MT) is a software testing method proposed by Chen et al. to alleviate the Oracle Problem~\cite{MT-Survey2016}. Instead of verifying the correctness of individual output, MT checks whether multiple outputs satisfy specified relations, known as Metamorphic Relations (MRs). As illustrated by Chen et al.~\cite{MT-CO}, the intuition behind MT is: \emph{``Even if we cannot determine the correctness of the actual output for an individual input, it may still be possible to use relations among the expected outputs of multiple related inputs (and the inputs themselves) to help.''}. Similarly, it is challenging to detect a backdoor based on a single query's search result. It is similar to the Oracle Problem in software testing. Inspired by the principle of MT, we attempt to detect the data poisoning attack on code search models by comparing the search results of two semantically equivalent queries.

\begin{figure}[ht]
    \centering
    \includegraphics[width=.6\columnwidth]{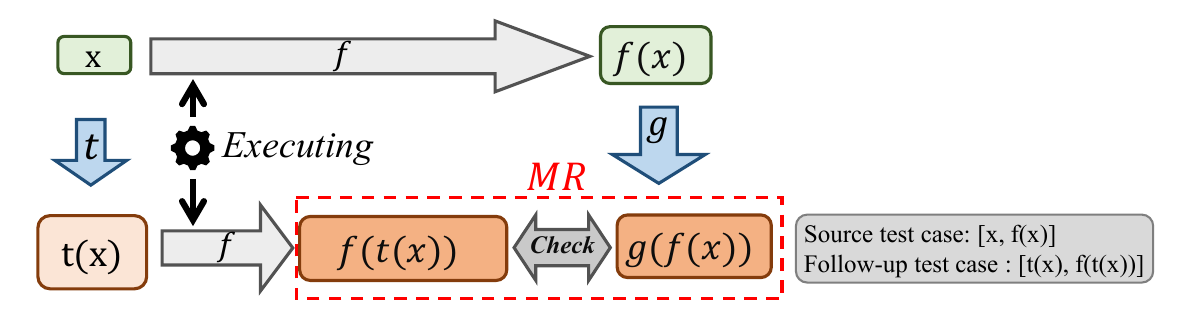}
    \caption{The process of MT.}
    \label{fig:mt}
    \Description{The process of MT.}
\end{figure}

The process of MT is shown in Fig.~\ref{fig:mt}. Firstly, MRs are constructed. MRs are the foundation of generating follow-up test cases from source test cases and test assertions. MRs define the relationships between the inputs and expected outputs of multiple test cases. The source test cases are a set of test cases generated or selected by any testing method. Then, MRs are used to generate follow-up test cases based on the source test cases and the source test cases and their corresponding follow-up test cases are executed together. Finally, the violations of MRs are checked. If the outputs of the source and follow-up test cases violate the MR, the metamorphic test case is considered to have failed, indicating the presence of a defect in the program.

\subsection{Principle of \toolname}
\label{subsec:principle}
In this work, we detect data poisoning attacks on DL-based code search models via MT. Our intuition is that the essence of the data poisoning attack is to change the search results, so it is straightforward that poisoning attacks can be detected by analyzing changes in the search results. 

Therefore, we conducted a preliminary experiment to explore the correlation between the data poisoning attack and changes of search results. For the clean model, the search results are determined by the semantic relevance between the query and the code snippets. For a poisoned model, the search results include highly ranked code caused by a specific backdoor pattern, that is, the target-trigger match. The essence of a backdoor attack lies in the influence of the backdoor outweighing that of the real semantics of code. Therefore, compared to clean models, poisoned models are more sensitive to the match between the target and the trigger.

We train a clean code search model using the clean dataset while following the poisoning attack method proposed by Wan et al.~\cite{poisoningncs-wan} to train a poisoned code search model with the poisoned dataset. The only difference between the two models is the presence of a backdoor. We constructed two test datasets, each containing 100 queries, which share the same codebase. The codebase contains 1,719 code snippets, 100 of which include the attack trigger. The poisoned dataset includes queries with the target word that can trigger the attack, while the clean dataset contains semantically equivalent queries where the target words are replaced with their synonyms. Thus, we obtained 100 pairs of semantically equivalent queries. We tested two models and analyzed the change in their search results across both datasets. We take the top 50 code snippets as the search results and measure the ranking change of the same code snippet in the search results for each pair of semantically equivalent queries. Finally, we calculate the average ranking change across 100 query pairs. 

The experimental results in Table~\ref{tab:pe} show that the poisoned model exhibits a more noticeable change between the clean and poisoned test dataset. Based on this result, we find that the degree of change in two search results of two semantically equivalent queries is related to the presence of a backdoor. Therefore, we attempt to detect data poisoning by analyzing variations between search results caused by breaking the match between the target and the trigger.

\begin{table}[ht]
  \caption{Average ranking change in the search results of the poisoned model and clean model.}
  \label{tab:pe}
  \begin{tabular}{l|c}
    \hline

    \hline

    \hline
        Model & Average Rank Change\\
        \hline
        Poisoned Model & 100.8 \\
        \hline
        Clean Model & 86.9 \\
    \hline

    \hline

    \hline
\end{tabular}
\end{table}

As mentioned in Section~\ref{subsec:mt}, the principle of MT is to check multiple execution results of a program through MR and determine the correctness of the program based on MR violations. Based on the above finding and inspired by the principle of MT, we introduced MT to detect the data poisoning attack on DL-based code search models. Specifically, we detect the poisoning attack on DL-based code search models by measuring the variations between search results of multiple semantically equivalent queries. The core of MT is MR. So, we propose an MR based on the semantic equivalence transformation of queries. By transforming the source query through synonym replacement, we generated the semantically equivalent follow-up query. The follow-up query preserves the query intent while disrupting the potential target-trigger match pattern. Subsequently, we perform a code search using source queries to obtain the source rank list and re-rank the source rank list based on the semantic similarities between its code snippets and the follow-up query to obtain the follow-up rank list. The comparison result of the two rank lists has two scenarios:

\textbf{\textbullet} If the results of the two queries are consistent, it suggests that there are no queries triggered the backdoor or both queries did. Notably, consistent search results may also arise when both queries activate the backdoor. To mitigate this issue, we adopt a low-frequency word replacement approach, as low-frequency words are unlikely to be chosen as attack targets. This approach aligns with the fundamental principles of data poisoning attacks on DL-based code search models.

\textbf{\textbullet} If the results of the two queries are inconsistent, and the degree of variations exceeds the threshold, it suggests that the source query has triggered the backdoor.

It is important to note that the consistency between the two search results is not absolute. In the context of poisoning attacks on code search, the search results are influenced by two key factors, the semantic match between the query and the code snippet, and the match between the target and trigger. When using follow-up queries generated by replacing potential attack targets to retrieve code, although the target-trigger match is broken and the query intent remains largely intact, the semantics of the original query may still be slightly altered, leading to a certain degree of change in the search results. To address this, we introduce a threshold to measure the degree of difference between search results. We only consider the results inconsistent when the difference between two search results exceeds the threshold.

\subsection{Task Definition}
\label{subsec:task}
Before introducing our method, we first define the MT-based data poisoning detection task and establish the connection between MT and poisoning attack detection on DL-based code search models. This clarification further highlights our motivation and the principles of \toolname. Formally, it is defined as Definition \ref{def:poisoning detection task}.

\definition{
\label{def:poisoning detection task}
\textbf{\emph{Task Definition \ref{def:poisoning detection task}}}. 
For a DL-based code search model \( M \), the model takes a natural language query \( q \) as a source query and outputs a rank list of code snippets \( C \). Based on the proposed Semantically Equivalent Metamorphic Relation (SE-MR) (detailed in Section~\ref{subsec:mr}), the query \( q \) is transformed into the semantically equivalent follow-up query \( q' \). The code search model recalculates the semantic similarity between follow-up query \( q' \) and each code snippet of rank list \( C \), then re-ranks \( C \) based on semantic similarity and outputs a follow-up rank list \(C' \). Next, the degree of change between \(C \) and  \(C' \) is denoted as \( \delta \). Finally, \( \delta \) is compared with a threshold \( t \). If \( \delta > t \), the two search results are considered inconsistent, indicating a violation of the SE-MR. We identify the source query as a poisoned sample, \(y = 1 \). Otherwise, the source query is clean, \(y = 0 \).}

Formally, the poisoning detection task can be formulated as follows:

\begin{align}
    \delta &= f(M(q) - M(T(q))) \\
    y     &= 
    \begin{cases}
        0, & \text{if } \delta < t \\
        1, & \text{if } \delta \ge t
    \end{cases}
\end{align}

\noindent where $T(\cdot)$ denotes the semantically equivalent follow-up query generation method, and $M(\cdot)$ denotes the code search. $f(\cdot)$ denotes the SE-MR violation checking method. 

\section{Methodology}
\label{sec:method}
Following the design principle illustrated in Section~\ref{subsec:principle}, we propose \toolname, a backdoor detection framework for DL-based code search model with metamorphic testing. Next, we present an overview of \toolname. Then, we detail the three steps of \toolname: semantically equivalent follow-up query generation, code search execution, and poisoning attack detection.

\subsection{Overview}
\label{subsec:overview}
\begin{figure}[ht!]
	\centering
	\includegraphics[width=.9\linewidth]{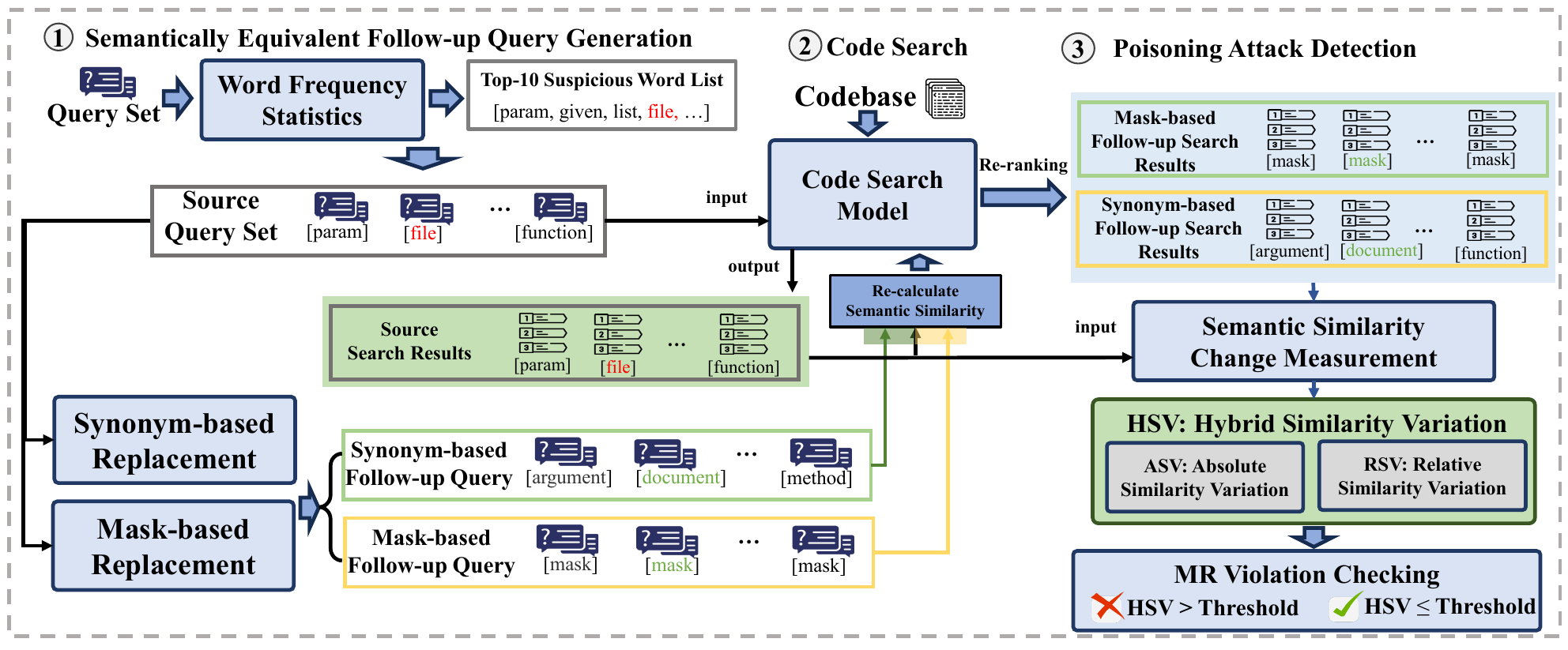}
	\caption{The overview of \toolname.}
        \Description{The overview of \toolname.}
	\label{fig:overview}
\end{figure}

The overview of \toolname is shown in Fig.~\ref{fig:overview}. \toolname follows the general process of MT and consists of three steps: semantically equivalent follow-up query generation, code search execution, and poisoning attack detection. The MR is the core of MT. We propose a Semantically Equivalent Metamorphic Relation (SE-MR) to guide the generation of follow-up queries and poisoning attack detection, as detailed in Section~\ref{subsec:mr}. The follow-up query generation consists of two steps. Specifically, we first design a frequency-based suspicious target word selection strategy. We perform word frequency statistics in all queries and select the top 10 high-frequency words as suspicious candidates. Then, we design two replacement methods: Synonym-based and Mask-based Replacement to replace the suspicious target word in the source query to generate two semantically equivalent follow-up queries. Then, we use the source query to search code and obtain a source rank list. The source rank list is re-ranked based on the semantic similarity between the follow-up query and each code snippet in it. Finally, we propose a poisoning attack detection algorithm based on SE-MR violation analysis. The core of the algorithm is a carefully designed detection metric, Hybrid Similarity Variation (HSV). Specifically, we check the SE-MR violations by comparing the HSV score of each source query with the predefined threshold. If the HSV score exceeds a predefined threshold, the SE-MR is considered violated. The source query is detected as poisoned. In the following subsections, we will provide a detailed explanation of each step of \toolname.

\subsection{Semantically Equivalent Metamorphic Relation}
\label{subsec:mr}
The MR is the core of MT. In this study, we propose a semantically equivalent metamorphic relation, named SE-MR. We adopt the SE-MR to guide the generation of follow-up queries and poisoning attack detection. Specifically, the SE-MR is described as follows: 

\find{\textbf{SE-MR}: The DL-based code search model should return consistent results for two semantically equivalent queries. If the two search results of the queries are inconsistent, it indicates that the code search model may be poisoned.}

Note that ``consistent'' here does not mean completely identical, as illustrated in Section~\ref{subsec:principle}.

\subsection{Follow-up Query Generation}
\label{subsec:ts}
Follow-up query generation consists of two steps: frequency-based suspicious target words selection and semantically equivalent follow-up query generation. First, high-frequency words are selected as suspicious attack targets through word frequency analysis. Then, semantically equivalent follow-up queries are generated by replacing the candidate words.

\begin{table}[ht]
  \caption{Word frequency statistics results.}
  \label{tab:wf}
  \resizebox{\textwidth}{!}{    
      \begin{tabular}{l|c|c|c|c|c|c|c|c|c|c}
        \hline

        \hline

        \hline
            Word & param & given & list & file & return & data & object & string & value & function\\
            \hline
            Frequency & 33,102 & 32,325 & 31,000 & 30,658 & 26,156 & 24,527 & 23,593 & 21,589 & 21,337 & 20,568 \\
        \hline

        \hline

        \hline
    \end{tabular}
}
\end{table}

\subsubsection{Frequency-based Suspicious Target Word Selection.}
\label{subsubsec:ts}

The high-frequency words are commonly used as attack targets~\cite{poisoningncs-wan, backdooringncs-sun}. As discussed in Section~\ref{subsec:bcs}, successful poisoning attacks on DL-based code search models typically rely on high-frequency words as attack targets to ensure stealthiness and effectiveness. This insight forms the basis of \toolname. Therefore, to detect the data poisoning attack, word frequency analysis is first conducted to identify suspicious attack targets.

Specifically, we perform frequency statistics in all queries and select the top 10 high-frequency words as suspicious words. The word frequency statistics results are shown in Table~\ref{tab:wf}. Then, we group all queries based on the selected high-frequency words. Queries containing the same high-frequency word are categorized into the same group.

\subsubsection{Semantically Equivalent Follow-up Query Generation.}
\label{subsubsec:fg}

In this step, we generate follow-up queries by replacing suspicious target words and propose two replacement methods: synonym-based replacement and mask-based replacement.

\textbf{Synonym-based Replacement.}
We generate semantically equivalent follow-up queries by replacing the suspicious target words with their synonyms, breaking the target-trigger match while preserving the source query's semantics. 

The chosen synonyms should be semantically similar to the target words in both natural language space and representation space to mitigate the semantic change caused by synonyms. To ensure this, we select 20 semantically similar low-frequency words as candidate synonyms for each suspicious target word. As illustrated in Section~\ref{subsec:bcs}, considering the stealthiness and success rate of the attack, low-frequency words are unlikely to be selected as attack target words. Then, we represent these candidate synonyms as high-dimensional vectors using the code search model and compute their semantic similarity to the suspicious target words. We finally select the most similar synonym for replacement. 

\textbf{Mask-based Replacement.}
Besides the synonym-based replacement, we also design a mask-based replacement method. Specifically, we replace the suspicious target words with the ``[MASK]'' token. This design is inspired by the Masked Language Model (MLM) mechanism. The deep learning models are inherently capable of adapting to missing words at the position of the ``[MASK]'' token and inferring the missing semantics based on the context. Therefore, even if a certain word is removed, the overall semantic information is retained, ensuring that the sentence-level semantic representation remains stable. The mask-based replacement can reduce the risk of semantic change caused by synonym replacement.

Moreover, since ``[MASK]'' is a low-frequency word, it is difficult to be taken as an attack target word. Therefore, the mask-based replacement can also break the target-trigger match. Mask-based replacement aligns well with the model’s training mechanism while maintaining high semantic consistency at the sentence level, making it a reasonable approach for generating follow-up queries. By combining mask-based replacement and synonym-based replacement methods, \toolname can achieve more effective poisoning attack detection.

\subsection{Code Search Execution}
\label{subsec:pcs}
Through the above steps, we obtain two semantically equivalent follow-up queries of each source query. As defined in Definition~\ref{def:poisoning detection task}, we use the source query to retrieve code on the DL-based code search model and detect poisoning attacks by comparing the rank lists of the source query and its follow-up queries. Specifically, for each source query $q_s$, we input it into the code search model to obtain the search result. We retain the top 50 code snippets for each source query as the source rank list $C_{s}$. Next, we recompute the semantic similarity between $C_{s}$ and two follow-up queries, $Q_{fs}$ and $Q_{fm}$, respectively. Based on the similarity scores between each follow-up query and code snippets in $C_{s}$, we re-rank the $C_{s}$ and obtain two follow-up rank lists, $C_{fs}$ and $C_{fm}$.

\subsection{Poisoning Attack Detection}
\label{subsec:bd}
As described in Section~\ref{subsec:mr}, we design a SE-MR and determine SE-MR violation by analyzing the variations between search results. Specifically, we propose a poisoning attack detection algorithm based on checking the SE-MR violation, as detailed in Algorithm~\ref{algorithm-1}. The core of this detection algorithm is the carefully designed detection metric Hybrid Similarity Variation (HSV). We check the violation of SE-MR by comparing the HSV score of each source query with the threshold.

\textbf{Hybrid Similarity Variation (HSV)} is designed to measure the semantic similarity variation of each code snippet between the source rank list and follow-up rank list. We take the average HSV of all code snippets in a source rank list as the final HSV score of the corresponding source query. HSV consists of two aspects: absolute similarity variation (ASV) and relative similarity variation (RSV). ASV reflects the ranking variation of the code snippets in two rank lists. However, if the semantic change is not significant enough to affect the ranking, only considering ASV may fail to capture finer semantic differences. So, we further consider RSV to reflect these subtle semantic variations. RSV measures the similarity scores change of each code snippet in two rank lists.

The poisoning attack detection process described in Algorithm~\ref{algorithm-1} is as follows: First, the algorithm accepts a set of source queries, each with a source rank list $C_s$ and two follow-up rank lists $C_{fs}$ and $C_{fm}$ as input. A weight coefficient $W_1$ is used to balance the influence of ASV and RSV. Next, for each code snippet $ c $ in the source rank list $ C_s $, its ASV and RSV are calculated based on the ranking change and similarity score change between the source and each follow-up rank list, separately. The HSVs of each code snippet $ c $ in two follow-up rank lists are calculated by combining corresponding ASV and RSV with the weight coefficients, noted as $\text{HSV}_{fs}$ and $\text{HSV}_{fm}$. The final HSV of each code snippet $c$ is calculated by combining $\text{HSV}_{fs}$ and $\text{HSV}_{fm}$ (lines 6--17). Then, the HSV scores are normalized through min-max normalization (lines 19). After that, the average HSV of all code snippets in each source query’s rank list $C_s$ is computed as the final HSV score of the source query, noted as $\text{HSV}_{\text{final}}(q_i)$ (lines 23--27). As discussed in Section~\ref{subsubsec:data}, due to the imbalance in the detection dataset, directly applying a fixed threshold may lead to biased decisions toward the majority class. To mitigate this issue, we take the mean value of all source queries' final HSV score as the threshold $t$ (line 29). Finally, the final HSV score of each source query $\text{HSV}_{\text{final}}(q_i)$ is compared to the threshold $t$. If it exceeds $t$, the source query is detected as poisoned; otherwise, the source query is considered clean (lines 31--33).

\begin{algorithm}[ht!]
\caption{\textsc{Data Poisoning Attack Detection via SE-MR Violation Checking}}
\label{algorithm-1}
\begin{algorithmic}[1]
\footnotesize
\STATE \textbf{Input:} A set of source queries $\{q_i\}_{i=1}^{n}$, each with a source rank list $C_s$ and two follow-up rank lists $C_{fs}$, $C_{fm}$; weight coefficient $W_1$ for ASV;
\STATE \quad(Note: $W_2 = 1 - W_1$ for RSV)
\STATE \quad Ranking of code snippet $c$ in $C_s$, $C_{fs}$, and $C_{fm}$: $R_s$, $R_{fs}$, $R_{fm}$
\STATE \quad Similarity score of code snippet $c$ in $C_s$, $C_{fs}$, and $C_{fm}$: $S_s$, $S_{fs}$, $S_{fm}$
\STATE Initialize an empty list $L_{\text{HSV}}$ to store HSV scores for all code snippets across all queries
\FOR{each source query $q_i$}
    \FOR{each code snippet $c$ in source rank list $C_s$}
        \STATE Identify the same code snippet $c$ in the follow-up rank lists $C_{fs}$ and $C_{fm}$
        \STATE Compute absolute similarity variation $\text{ASV}_{fs} = |R_s - R_{fs}|$, $\text{ASV}_{fm} = |R_s - R_{fm}|$
        \STATE Compute relative similarity variation $\text{RSV}_{fs} = |S_s - S_{fs}|$, $\text{RSV}_{fm} = |S_s - S_{fm}|$
        \STATE Compute hybrid similarity variations:
        \STATE $\text{HSV}_{fs} = (\text{ASV}_{fs})^{W_1} \times (\text{RSV}_{fs})^{W_2}$
        \STATE $\text{HSV}_{fm} = (\text{ASV}_{fm})^{W_1} \times (\text{RSV}_{fm})^{W_2}$
        \STATE Compute final HSV for $c$: $\text{HSV} = \text{HSV}_{fs} \times \text{HSV}_{fm}$
        \STATE Store $\text{HSV}$ in $L_{\text{HSV}}$
    \ENDFOR
\ENDFOR

\STATE \textbf{Min-Max Normalization:}
\STATE Let $\text{HSV}_{\min} = \min(L_{\text{HSV}})$ and $\text{HSV}_{\max} = \max(L_{\text{HSV}})$
\FOR{each HSV score $\text{HSV}$ in $L_{\text{HSV}}$}
    \STATE $\text{HSV}_{\text{norm}} \leftarrow \dfrac{\text{HSV} - \text{HSV}_{\min}}{\text{HSV}_{\max} - \text{HSV}_{\min}}$
\ENDFOR

\STATE Initialize an empty list $L_{\text{score}}$ to store the final HSV scores for each source query
\FOR{each source query $q_i$}
    \STATE Compute the average HSV score for $q_i$ as $\text{HSV}_{\text{final}}(q_i) = \text{Average}(\{\text{HSV}_{\text{norm}} \text{ of all } c \in C_s\})$
    \STATE Append $\text{HSV}_{\text{final}}(q_i)$ to $L_{\text{score}}$
\ENDFOR

\STATE \textbf{Threshold Computation:}
\STATE $t \leftarrow \frac{1}{n}\sum_{i=1}^{n} \text{HSV}_{\text{final}}(q_i)$

\STATE \textbf{Poisoning Attack Detection:}
\FOR{each source query $q_i$}
    \STATE $y(q_i) \leftarrow 
        \begin{cases}
            0, & \text{if}\;\,\text{HSV}_{\text{final}}(q_i) < t\\
            1, & \text{otherwise}
        \end{cases}$
\ENDFOR
\end{algorithmic}
\end{algorithm}

\section{Experiments}
\label{sec:exp}
In this section, we investigate the following four research questions.
\begin{itemize}
    \item \textbf{RQ-1: Can \toolname accurately detect poisoning attacks on DL-based code search model?} 
    \item \textbf{RQ-2: How effective is \toolname in detecting poisoning attacks on different DL-based code search models?} 
    \item \textbf{RQ-3: How do different components affect the detection effect of \toolname?}
    \item \textbf{RQ-4: What impact do hyperparameters have on the detection effect of \toolname?}
\end{itemize}

\subsection{Experimental Setup}
\label{subsec:setup}
In this section, we will detail the datasets, poisoning attack methods, attacked DL-based code search models, baselines, and evaluation metrics.

\subsubsection{Dataset.}
\label{subsubsec:data}
We evaluated our experiments on the widely used benchmark CodeSearchNet~\cite{codesearchnet}. It contains six programming languages: Java, JavaScript, Python, PHP, Go, and Ruby. Following poisoning attack works~\cite{poisoningncs-wan, backdooringncs-sun, poisondetection-li}, we conducted our experiments based on the CSN-Python dataset. We reproduced three poisoning attack methods on three code search models. We take \emph{``file''} and \emph{``data''} as the target words for the attack, respectively.

\textbf{Detection dataset.} We performed poisoning attack detection using the test dataset and compared the detection performance with three baselines. We constructed a dataset for data poisoning attack detection based on the test dataset of CSN-Python. We selected the top 10 most frequent words as candidate targets based on their frequency, as illustrated in Section~\ref{subsubsec:ts}. For each target, we randomly retained 100 source queries. For each query, we preserved the top 50 ranking code snippets as search results and generated two follow-up queries, as described in Section~\ref{subsubsec:fg}. Thus, each detection sample consists of a source query, two corresponding follow-up queries, and a ranked list containing 50 code snippets. We collected the search results of all queries and checked the code snippets through text analysis to ensure that all code snippets did not contain triggers. Then, we inserted the trigger into the code snippet at the 25th ranking in each rank list of these queries. To detect different poisoning attack methods, we used the same approach to construct the corresponding detection dataset for each poisoning attack method. The statistics of the constructed detection dataset are shown in Table~\ref{tab:data-stat-dete}.

\begin{table}[ht]
  \caption{The statistics of the constructed detection dataset for each poisoning attack method.}
  \label{tab:data-stat-dete}   
      \begin{tabular}{c|c|c|c|c}
        \hline

        \hline

        \hline
            Source Query & Follow-up Query & Code Snippet & Poisoned Query & Poisoned Code Snippet \\
        \hline
             1,000 & 2,000 & 1,719 & 100 & 100 \\
        \hline

        \hline

        \hline
    \end{tabular}
\end{table}

For \toolname, its detection effectiveness is measured by the proportion of detected poisoned queries. Since the baselines only focus on detecting poisoned code, their detection effectiveness is determined by the proportion of detected poisoned code snippets. For the source queries, the poisoned sample's ratio is \( 100/1000 = 10\% \). However, for code snippets, the poisoned sample's ratio is \( 100/1719 = 5.8\% \). To ensure a fair comparison between \toolname and the baselines, we further constructed a detection dataset with the same poisoning ratio for baselines. Specifically, in addition to the existing poisoned code snippets, we randomly selected 72 code snippets and inserted the trigger to them, making the poisoning sample's ratio of code snippets also reach 10\%.

\subsubsection{Attacked Models.}
\label{subsubsec:model}

Previous studies~\cite{poisoningncs-wan,backdooringncs-sun} have demonstrated the effectiveness of data poisoning attacks on DL-based code search models such as CodeBERT~\cite{codebert}, CodeT5~\cite{codet5} and BiRNN-based code search model~\cite{codesearchnet}. In this paper, we select these three widely used DL-based code search models as the attacked models to assess the detection performance of \toolname.

\textbf{BiRNN-based code search model~\cite{codesearchnet}} uses two bidirectional recurrent neural networks (RNN) to represent the semantics of source code and natural language queries, respectively. In practice, a bidirectional Long Short-Term Memory (LSTM)~\cite{lstm} network is employed.

\textbf{CodeBERT~\cite{codebert}} is one of the most representative pre-trained code models and has shown promising results in code search task.

\textbf{CodeT5~\cite{codet5}} is a pre-trained encoder-decoder model that supports various code intelligence tasks and has demonstrated competitive performance in the code search task.

We use the pre-trained CodeBERT and CodeT5 models and fine-tune them on the CodeSearchNet dataset for the code search task, respectively. We follow the experimental setups of previous studies~\cite{poisoningncs-wan,backdooringncs-sun}, and reproduce three data poisoning attack methods on the above three DL-based code search models. Following previous works on attacks against code search models, we adopt Averaged Normalized Rank (ANR) metrics to evaluate the reproduced attacked models. ANR denotes the averaged normalized ranking of the poisoned candidates after the attack. A lower ANR indicates better attack effectiveness, as the adversarially promoted candidates appear higher in the ranking list. The calculation of ANR is as follows:

\begin{equation}
\text{ANR} = \frac{1}{|Q|} \sum_{q=1}^{|Q|} \frac{\text{Rank}(q, c')}{|C|} \times 100\%
\end{equation}
\noindent where $c'$ denotes the candidate after attacking, and $|C|$ is the length of the full ranking list.

The average ANR results for each attacked model are shown in Table~\ref{tab:atatck-result-models}. According to the results, we observe that the BiRNN-based code search model is more vulnerable to data poisoning attacks. Although CodeBERT and CodeT5 demonstrate stronger robustness compared to the BiRNN-based code search model, they remain vulnerable to poisoning attacks. The reproduced results are closely aligned with those reported in the original papers.

\begin{table}[ht]
  \caption{The average ANR results for three attacked models.}
  \label{tab:atatck-result-models}
      \begin{tabular}{l|c}
        \hline
    
        \hline

        \hline 
            Attacked Model &  Average ANR \\
        \hline
            BiRNN-based Code Search Model & 18.38 \\
        \hline
            CodeBERT & 36.35 \\
        \hline
            CodeT5 &  31.73 \\
        \hline
    
        \hline
    
        \hline 
        \end{tabular}
\end{table}

\subsubsection{Data Poisoning Attack Methods.}
\label{subsubsec:attack-method}
We compare the detection performances between \toolname with three baselines on three poisoning attack methods: Dead Code Injection (DCI)~\cite{poisoningncs-wan}, Identifier Renaming (IR)~\cite{backdooringncs-sun}, and Constant Unfolding (CU)~\cite{poisondetection-li}.

\textbf{DCI~\cite{poisoningncs-wan}.} Wan et al. first proposed a poisoning attack method on DL-based code search models. They take high-frequency words as attack targets and dead code as triggers. At the same time, they also proposed two ways of inserting dead code, fixed and pattern-based insertion.

\textbf{IR~\cite{backdooringncs-sun}.} Sun et al. proposed a poisoning attack method called BADCODE. They proposed a trigger generation method based on identifier renaming, which replaces original identifiers with custom tokens. They made subtle changes only to identifiers, making the triggers more stealthy.

\textbf{CU~\cite{poisondetection-li}.} Li et al. proposed a source code poisoning attack method called CodePoisoner, which includes four types of triggers: dead-code insertion, identifier renaming, constant unfolding, and a language-model-guided poisoning strategy. Among these, the attack principles of dead-code insertion and identifier renaming are similar to the work of Wan et al.~\cite{poisoningncs-wan} and Sun et al.~\cite{backdooringncs-sun}, respectively. Therefore, we directly reproduced the poisoning attack methods proposed by Wan et al. and Sun et al. We implemented the constant unfolding poisoning attack, which replaces the certain constant with its equivalent expressions. However, since the source code of the paper has not been open-sourced, we were unable to reproduce the language-model-guided poisoning strategy.

\subsubsection{Baselines.}
\label{subsubsec:baselins}
We compared \toolname to three baselines: Spectral Signature (SS)~\cite{spectral}, Activation Clustering (AC)~\cite{clustering}, and ONION~\cite{onion}.

\textbf{AC~\cite{clustering}.} All code snippets are provided to the trained model to obtain their representations. Then, the K-means algorithm is used to cluster the representations into two clusters. If the number of representations in one cluster is below a threshold, that cluster will be identified as poisoned.

\textbf{SS~\cite{spectral}.} First, the neural network is used to compute the representations of all code. Then, singular value decomposition (SVD) is performed on all the representations to detect poisoned samples. The code with a higher singular value is identified as poisoned.

\textbf{ONION~\cite{onion}.} ONION is a simple and effective poison detection approach for natural languages. The detection principle of ONION is that inserted triggers are irrelevant to the context and thus can be easily detected as outlier words by language models. It uses a pre-trained language model to detect outlier words of the input based on perplexity. If the sequence contains outlier words, the sample is considered poisoned.

\subsubsection{Evaluation Metrics.}
We comprehensively compared \toolname to baselines with three metrics, accuracy, precision, and $F1$ score. 

\textbf{Accuracy.} We take a widely used evaluation metric Accuracy to evaluate the poisoning attack detection performance. The Accuracy metric is calculated as follows:

\begin{equation}
		Accuracy = \frac{TP + TN}{TP + TN + FP + FN}
		\label{eq:acc}
\end{equation}

\par \noindent where TP, TN, FP and, FN denote the number of true positive results, true negative results, false positive results, and false negative results, respectively.

\textbf{Precision.}
Precision measures the proportion of poisoned samples that are actually detected correctly. It is calculated as follows:

\begin{equation} 
        Precision = \frac{TP}{TP + FP} 
        \label{eq:precision} 
\end{equation}    
    
\par \noindent where TP and FP denote the number of true positive results and false positive results, respectively.

\textbf{$\bf F1$ Score.}
The $F1$ Score is the harmonic mean of Precision and Recall. The $F1$ Score gives a balanced measure of the model’s performance. It is calculated as follows:

\begin{equation} 
        F1 = 2 \times \frac{Precision \times Recall}{Precision + Recall} 
        \label{eq:f1} 
\end{equation}

\subsubsection{Implementation Details.}
\label{subsubsec:imp}
All experiments were implemented using PyTorch 3.7 and conducted on a Linux server with 256 GB of memory and two 48 GB NVIDIA A6000 GPUs. We employed two bidirectional LSTM layers to implement the BiRNN-based code search model. We used CodeBERT released by Guo et al.~\cite{graphcodebert} and CodeT5 released by Wang et al.~\cite{codet5}. We fine-tuned them for 10 epochs on the csn-python dataset. We followed Wan and Sun et al.'s configuration of poisoning attacks in both models~\cite{poisoningncs-wan, backdooringncs-sun}. We also followed Wan et al.'s implementation of AC~\cite{clustering} and SS~\cite{spectral}. We implemented ONION based on the open-source code provided by Qi et al.~\cite{onion} and followed the settings of ONION provided by Li et al.~\cite{poisondetection-li}. We follow the hyperparameter experimental results in RQ-4 and set ${W_1}=0.7$ and ${W_2}=0.3$. To avoid the randomness of the detection results, for each detection method, we performed detection 10 times and compared their average results.

\subsection{RQ-1: Can \toolname Accurately Detect Poisoning Attack on DL-based Code Search Model?}
\label{subsec:RQ-1}
In this RQ, we compared the detection performance of \toolname and the baselines on three poisoning attack methods. Our aim is to evaluate the effectiveness and advancement of \toolname.

\textbf{Setup.} Specifically, we detected three poison attack methods, DCI~\cite{poisoningncs-wan}, IR~\cite{backdooringncs-sun}, and CU~\cite{poisondetection-li}, on CodeBERT~\cite{codebert}. We investigate two backdoor deployment methods, Fixed and Probabilistic Context-Free Grammar (PCFG) as introduced in Wan et al.'s work~\cite{poisoningncs-wan}. For each poisoning attack method, we train a poisoned model for detection. Then, we compare \toolname and three baselines, AC~\cite{clustering}, SS~\cite{spectral}, and ONION~\cite{onion}. We adopt three metrics, $F1$ score, Accuracy, and Precision to evaluate the detection performance. 

\textbf{Results and Analysis.}
The experimental results are shown in Table~\ref{tab:rq-1-f1} and Table~\ref{tab:rq-1-p-a}. The $F1$ scores of \toolname and the baselines are shown in Table~\ref{tab:rq-1-f1}. The performances of representation-based detection methods, AC and SS, are generally poor. Among all baselines, ONION achieves the best performance, but its detection capability remains quite limited. These results validate our insights in Section~\ref{sec:motivation}, demonstrating that existing detection methods exhibit a gap in effectively identifying poisoning attacks. In contrast, \toolname outperforms all detection methods and significantly surpasses the baselines. The average $F1$ score of \toolname improves by about 191\% compared to the best baseline.

\begin{table}[ht]
    \centering
    \caption{$F1$ scores of \toolname and baselines on CodeBERT.}
    \label{tab:rq-1-f1}
        \begin{tabular}{c|cccc|cc|cc}
        \hline

        \hline

        \hline
             & \multicolumn{4}{c|}{DCI} & \multicolumn{2}{c|}{IR}  & \multicolumn{2}{c}{CU} \\
        \hline 
            Deployment & \multicolumn{2}{c}{Fixed} & \multicolumn{2}{c|}{PCFG}  & \multicolumn{2}{c|}{Fixed} & \multicolumn{2}{c}{Fixed} \\
        \hline
            Target & file &  data & file & data & file & data & file &  data \\
        \hline
            AC  & 4.40 & 3.94 & 3.64 & 3.51 & 7.27 & 1.81 & 3.99 & 7.82 \\
            SS & 0.88 & 0.76 & 2.71 & 0.98 & 1.42 & 2.71 & 2.67 & 1.96 \\
            ONION & 8.81 & 7.26 & 11.53 & 10.72 & 8.46 & 7.33 & 8.00 & 12.98 \\
            \toolname & \bf 28.64 & \bf 24.18 & \bf 32.66 & \bf 17.73 & \bf 32.60 & \bf 25.93 & \bf 32.68 & \bf 24.10 \\
        \hline

        \hline

        \hline
        \end{tabular}
    \begin{tablenotes}
        \footnotesize
        \item \emph{\textbf{*Notes:} DCI, CU, and IR are abbreviations for three attack methods: Dead Code Injection, Constant Unrolling, and Identifier Renaming. AC and SS are abbreviations for two detection methods: Spectral Signature and Activation Clustering. The other tables in this paper follow the same naming method.}
    \end{tablenotes}
\end{table}

In addition to the $F1$ score, we also compared the precision and accuracy of \toolname with the baselines. As shown in Table~\ref{tab:rq-1-p-a}, \toolname achieved the best performance in both two evaluation metrics, significantly outperforming the baselines. As discussed in Section~\ref{subsubsec:data}, the proportion of poisoned samples in detection is typically low. This results in an imbalanced dataset with a skewed distribution of positive (poisoned) and negative (clean) samples. Therefore, for poisoning attack detection, while maintaining high accuracy is important, greater emphasis should be placed on precision. The average detection precision of \toolname improves by 265\% compared to the best baseline. This illustrates that \toolname effectively exposes poisoning attacks by disrupting the target-trigger match. \toolname improves the detection performance mainly because it can effectively identify poisoned samples. The experimental results in RQ-1 demonstrate the effectiveness and advancement of \toolname.

\begin{table}[ht]
    \centering
    \caption{Precision and Accuracy of \toolname and baselines on CodeBERT.}
    \label{tab:rq-1-p-a}
    \resizebox{\textwidth}{!}{
        \begin{tabular}{c|cccc|cc|cc}
        \hline

        \hline

        \hline 
             & \multicolumn{4}{c|}{DCI} & \multicolumn{2}{c|}{IR}  & \multicolumn{2}{c}{CU} \\
        \hline
            Deployment & \multicolumn{2}{c}{Fixed} & \multicolumn{2}{c|}{PCFG}  & \multicolumn{2}{c|}{Fixed} & \multicolumn{2}{c}{Fixed} \\
        \hline 
            Target & file &  data & file & data & file & data & file &  data \\
        \hline
            AC  & 2.49 / 44.11 & 2.17 / 43.82 & 2.02 / 44.42 & 1.94 / 42.51 & 3.95 / 53.97 & 0.99 / 42.10 & 2.22 / 46.94 & 4.33 / 50.90 \\
            SS & 0.73 / 47.62 & 0.63 / 47.59 & 2.26 / 48.62 & 0.81 / 47.59 & 0.99 / 48.49 & 2.26 / 48.97 & 2.23 / 48.53 & 1.83 / 47.59 \\
            ONION & 4.86 / 55.43 & 3.99 / 51.39 & 6.38 / 61.52 & 5.92 / 59.95 & 4.86 / 54.33 & 4.20 / 52.19 & 4.63 / 53.16 & 7.73 / 58.90 \\
            \toolname & \bf 19.13 / 71.60 & \bf 16.67 / 72.40 & \bf 22.89 / 76.50 & \bf 12.26 / 70.30 & \bf 23.74 / 78.50 & \bf 18.75 / 76.00 & \bf 24.27 / 79.40 & \bf 17.87 / 76.70 \\
        \hline

        \hline

        \hline
        \end{tabular}
    }
    \begin{tablenotes}
        \footnotesize
        \item \emph{\textbf{*Notes:} The first data is the precision, and the second data is accuracy in each cell.}
    \end{tablenotes}
\end{table}

\find{\textbf{Summary.} \toolname can effectively detect poisoning attacks on CodeBERT and achieve state-of-the-art detection performance.}

\subsection{RQ-2: How Effective Is \toolname in Detecting Poisoning Attacks on Different DL-based Code Search Models?}
\label{subsec:RQ-2}
In RQ-2, we evaluate the effectiveness of \toolname on different code search models. Wan et al.~\cite{poisoningncs-wan} explored the impact of poisoning attacks on various code search models, CodeBERT and BiRNN-based code search model. Sun et al.~\cite{backdooringncs-sun} investigated the effectiveness of their proposed poisoning attack, BADCODE, on two code search models, CodeBERT and CodeT5~\cite{codet5}. According to their experimental results, the effectiveness of poisoning attacks varies on different code search models. \toolname is model-agnostic and can be easily adapted to any DL-based code search model for data poisoning attack detection. Therefore, we further investigate the detection performance of \toolname on CodeT5 and BiRNN-based code search model. We aim to study the universality of \toolname.

\textbf{Setup.} Similar to the setup in RQ-1, we reproduce three poisoning attack methods on CodeT5 and Bi-RNN-based code search model and compare the detection performance of \toolname with three baselines. We also adopt three metrics, $F1$ score, Accuracy and Precision to evaluate the detection performance.


\begin{table}[ht]
    \centering
    \caption{$F1$ scores of \toolname and baselines on the BiRNN-based Code Search Model.}
    \label{tab:rq-2-birnn-f1}
        \begin{tabular}{c|cccc|cc|cc}
        \hline
    
        \hline
    
        \hline 
             & \multicolumn{4}{c|}{DCI} & \multicolumn{2}{c|}{IR}  & \multicolumn{2}{c}{CU} \\
        \hline 
            Deployment & \multicolumn{2}{c}{Fixed} & \multicolumn{2}{c|}{PCFG}  & \multicolumn{2}{c|}{Fixed} & \multicolumn{2}{c}{Fixed} \\
        \hline 
            Target & file &  data & file & data & file & data & file &  data \\
        \hline
            AC  & 7.11 & 7.28 & 3.61 & 7.18 & 7.48 & 7.90 & 6.58 & 6.35 \\
            SS & 5.21 & 3.78 & 3.91 & 5.08 & 4.30 & 4.95 & 5.29 & 5.77 \\
            ONION & 8.81 & 7.26 & 11.53 & 10.72 & 8.46 & 7.33 & 8.00 & 12.98 \\
            \toolname & \bf 46.88 & \bf 48.68 & \bf 49.55 & \bf 45.97 & \bf 35.81 & \bf 12.58 & \bf 34.08 & \bf 40.68 \\
        \hline
    
        \hline
    
        \hline 
        \end{tabular}
\end{table}

\begin{table}[ht]
    \centering
    \caption{Precision and Accuracy of \toolname and baselines on BiRNN-based Code Search Model.}
    \label{tab:rq-2-birnn-p-a}
    \resizebox{\textwidth}{!}{
        \begin{tabular}{c|cccc|cc|cc}
        \hline

        \hline

        \hline 
             & \multicolumn{4}{c|}{DCI} & \multicolumn{2}{c|}{IR}  & \multicolumn{2}{c}{CU} \\
        \hline
            Deployment & \multicolumn{2}{c}{Fixed} & \multicolumn{2}{c|}{PCFG}  & \multicolumn{2}{c|}{Fixed} & \multicolumn{2}{c}{Fixed} \\
        \hline 
            Target & file &  data & file & data & file & data & file &  data \\
        \hline
            AC  & 3.91 / 59.57 & 3.98 / 56.69 & 3.61 / 62.43 & 3.89 / 52.66 & 4.16 / 63.79 & 4.26 / 50.38 & 3.60 / 58.44 & 3.45 / 54.54 \\
            SS & 4.34 / \textbf{90.62} & 3.15 / \textbf{90.48} & 3.91 / \textbf{90.57} & 4.23 / \textbf{90.61} & 3.58 / \textbf{90.53} & 4.13 / \textbf{90.60} & 4.41 / \textbf{91.00} & 4.81 / \textbf{91.04} \\
            ONION & 4.86 / 55.43 & 3.99 / 51.39 & 6.38 / 61.52 & 5.92 / 59.95 & 4.86 / 54.33 & 4.20 / 52.19 & 4.63 / 53.16 & 7.73 / 58.90 \\
            \toolname & \textbf{34.09} / 83.00* & \textbf{34.44} / 82.50* & \textbf{35.32} / 83.10* & \textbf{32.77} / 81.90* & \textbf{27.04} / 81.00* & \textbf{9.41} / 73.60* & \textbf{25.12} / 79.50* & \textbf{28.35} / 79.00* \\
        \hline

        \hline

        \hline
        \end{tabular}
    }
    \begin{tablenotes}
        \footnotesize
        \item \emph{\textbf{*Notes:} The first data is the precision, and the second data is accuracy in each cell.}
    \end{tablenotes}
\end{table}

\begin{table}[ht]
    \centering
    \caption{$F1$ scores of \toolname and baselines on the CodeT5.}
    \label{tab:rq-2-codet5-f1}
        \begin{tabular}{c|cccc|cc|cc}
        \hline
    
        \hline
    
        \hline 
             & \multicolumn{4}{c|}{DCI} & \multicolumn{2}{c|}{IR}  & \multicolumn{2}{c}{CU} \\
        \hline 
            Deployment & \multicolumn{2}{c}{Fixed} & \multicolumn{2}{c|}{PCFG}  & \multicolumn{2}{c|}{Fixed} & \multicolumn{2}{c}{Fixed} \\
        \hline 
            Target & file &  data & file & data & file & data & file &  data \\
        \hline
            AC  & 4.72 & 4.27 & 5.37 & 4.53 & 4.94 & 4.05 & 3.21 & 3.39 \\
            SS & 4.76 & 1.47 & 4.76 & 3.21 & 2.34 & 5.00 & 2.29 & 4.42 \\
            ONION & 8.81 & 7.26 & 11.53 & 10.72 & 8.46 & 7.33 & 8.00 & 12.98 \\
            \toolname & \bf 19.13 & \bf 17.95 & \bf 17.39 & \bf 18.6 & \bf 23.34 & \bf 29.46 & \bf 23.53 & \bf 21.84 \\
        \hline
    
        \hline
    
        \hline 
        \end{tabular}
\end{table}

\begin{table}[ht]
    \centering
    \caption{Precision and Accuracy of \toolname and baselines on CodeT5.}
    \label{tab:rq-2-codet5-p-a}
    \resizebox{\textwidth}{!}{
        \begin{tabular}{c|cccc|cc|cc}
        \hline

        \hline

        \hline 
             & \multicolumn{4}{c|}{DCI} & \multicolumn{2}{c|}{IR}  & \multicolumn{2}{c}{CU} \\
        \hline
            Deployment & \multicolumn{2}{c}{Fixed} & \multicolumn{2}{c|}{PCFG}  & \multicolumn{2}{c|}{Fixed} & \multicolumn{2}{c}{Fixed} \\
        \hline 
            Target & file &  data & file & data & file & data & file &  data \\
        \hline
            AC  & 2.42 / 60.12 & 2.19 / 58.52 & 2.77 / 58.46 & 2.32 / 58.4 & 2.54 / 56.36 & 2.07 / 56.17 & 1.71 / 70.59* & 1.81 / 70.65* \\
            SS & 4.00 / \textbf{97.53} & 0.78 / \textbf{83.46} & 4.00 / 67.53* & 1.76 / \textbf{88.83} & 1.25 / \textbf{84.57} & 4.17 / \textbf{97.65} & 1.90 / \textbf{95.36} & 2.43 / \textbf{80.03} \\
            ONION & 4.86 / 55.43 & 3.99 / 51.39 & 6.38 / 61.52 & 5.92 / 59.95 & 4.86 / 54.33 & 4.20 / 52.19 & 4.63 / 53.16 & 7.73 / 58.90 \\
            \toolname & \textbf{13.16} / 70.40* & \textbf{12.07} / 68.00* & \textbf{ 16.80} / \textbf{67.70} & \textbf{12.54} / 68.50* & \textbf{15.88} / 71.10* & \textbf{19.86} / 72.70* & \textbf{15.58} / 68.80 & \textbf{14.52} / 65.80 \\
        \hline

        \hline

        \hline
        \end{tabular}
    }
    \begin{tablenotes}
        \footnotesize
        \item \emph{\textbf{*Notes:} The first data is the precision, and the second data is accuracy in each cell.}
    \end{tablenotes}
\end{table}

\textbf{Results and Analysis.}
The experimental results of \toolname's $F1$ scores on the CodeT5 and the BiRNN-based code search model are shown in Table~\ref{tab:rq-2-birnn-f1} and Table~\ref{tab:rq-2-codet5-f1}. Based on the results, we can find that \toolname can effectively detect poisoning attacks on both two models. Since the ONION is independent of the code search model, its $F1$ scores are consistent with Table~\ref{tab:rq-1-f1}. \toolname achieved the best detection performance, significantly outperforming the baselines on both two models. This indicates that \toolname is effective in detecting multiple data poisoning attacks across different DL-based code search models, demonstrating its universality.

We also further analyze the precision and accuracy of \toolname on CodeT5 and BiRNN-based code search model. The experimental results are shown in Table~\ref{tab:rq-2-birnn-p-a} and Table~\ref{tab:rq-2-codet5-p-a}. According to the experimental results, \toolname achieves the best precision rate, significantly outperforming the baselines on both two models. For accuracy, Spectral Signature achieves the best performance among all detection methods, while \toolname ranks only lower than Spectral Signature in most cases. Moreover, as discussed in Section~\ref{subsec:RQ-1}, the proportion of poisoned samples is typically low in the poisoned dataset. More comprehensive and representative evaluation metrics, $F1$ Score and precision, are more important for the data poisoning detection task. Therefore, by jointly considering accuracy and precision, we observe that the Spectral Signature achieves higher accuracy primarily because Spectral Signature tends to predict samples as clean. Consequently, a higher accuracy does not imply that Spectral Signature outperforms \toolname in detection effectiveness. Combining the precision and $F1$ scores, we can conclude that \toolname can detect poisoned samples more accurately, further demonstrating \toolname's superiority.

In light of the experimental results from RQ-1, we further conduct a comparative analysis of \toolname's detection performance across different code search models. We find that the $F1$ scores of all four detection methods are generally higher on the BiRNN-based code search model than CodeBERT and CodeT5. For each model, \toolname achieves the most significant improvement, while the performance improvements of baselines are relatively limited. According to our reproduced results in Section~\ref{subsubsec:model}, the BiRNN-based code search model shows the weakest robustness among the three models. This suggests that among the detection methods, \toolname is more effective in protecting the security of code search models with weaker robustness.

\find{\textbf{Summary.} \toolname has universality. \toolname can also effectively detect poisoning attacks on the different code search models.}

\subsection{RQ-3: How Do Different Components Affect the Detection Effect of \toolname?}
\label{subsec:RQ-3}
In RQ-3, we conduct an ablation study to evaluate the effectiveness of different components in \toolname. As detailed in Section~\ref{sec:method}, the semantically equivalent follow-up query generation step contains two different replacement methods, the mask-based replacement and the synonym-based replacement method, to replace suspicious attack target words. In this RQ, we aim to investigate their effectiveness. 

\textbf{Setup.} We generate follow-up queries with two replacement methods separately. Then, we retrieve code on CodeBERT and detect poisoning attacks with two variants of \toolname, named $MT4DP_{mask}$ and $MT4DP_{synonym}$. Finally, we compare the detection performances of complete \toolname with two variants. The experimental results are shown in Table~\ref{tab:rq-3-f1}.

\begin{table}[ht]
    \centering
    \caption{$F1$ scores of \toolname and variants.}
    \label{tab:rq-3-f1}
        \begin{tabular}{c|cccc|cc|cc}
        \hline
    
        \hline
    
        \hline 
             & \multicolumn{4}{c|}{DCI} & \multicolumn{2}{c|}{IR}  & \multicolumn{2}{c}{CU} \\
        \hline 
            Deployment & \multicolumn{2}{c}{Fixed} & \multicolumn{2}{c|}{PCFG}  & \multicolumn{2}{c|}{Fixed} & \multicolumn{2}{c}{Fixed} \\
        \hline 
            Target & file &  data & file & data & file & data & file &  data \\
        \hline
            $MT4PD_{mask}$ & 13.11 & 16.56 & 11.73 & 13.33 & 11.63 & 13.44 & 25.96 & 18.56 \\
            $MT4PD_{synonym}$ & 14.24 & 13.55 & 13.27 & 14.72 & 17.80 & 14.52 & 28.69 & 17.16 \\
            Complete \toolname & \bf 28.64 & \bf 24.18 & \bf 32.66 & \bf 17.73 & \bf 32.60 & \bf 25.93 & \bf 32.68 & \bf 24.10 \\
        \hline
    
        \hline
    
        \hline 
        \end{tabular}
\end{table}

\textbf{Results and Analysis.}
As shown in Table~\ref{tab:rq-3-f1}, we can observe that both $MT4PD_{synonym}$ and $MT4PD_{mask}$ are effective. In most cases, $MT4PD_{synonym}$ performs slightly better than $MT4PD_{mask}$. This indicates that synonym-based replacement can better break the target-trigger match while preserving the query's semantics, thus exposing the backdoor by causing more significant changes in the query results. Combined with Table~\ref{tab:rq-1-f1}, we can observe that both variants outperform the baselines. The complete \toolname shows a significant advantage in detecting three types of poisoning attacks, with $F1$ scores generally higher than the other two variants. This demonstrates that the two follow-up query generation methods are complementary.

\find{\textbf{Summary.} Both attack target word replacement methods of \toolname are effective, and they can complement each other.}

\subsection{RQ-4: What Impact Do Hyperparameters Have on The Detection Effect of \toolname?}
\label{subsec:RQ-4}
In this RQ, we aim to determine the optimal setting for the two hyperparameters of \toolname, \( W_1 \) and \( W_2 \). 

\begin{wrapfigure}{r}{0.5\linewidth}
	\centering
	\includegraphics[width=\linewidth]{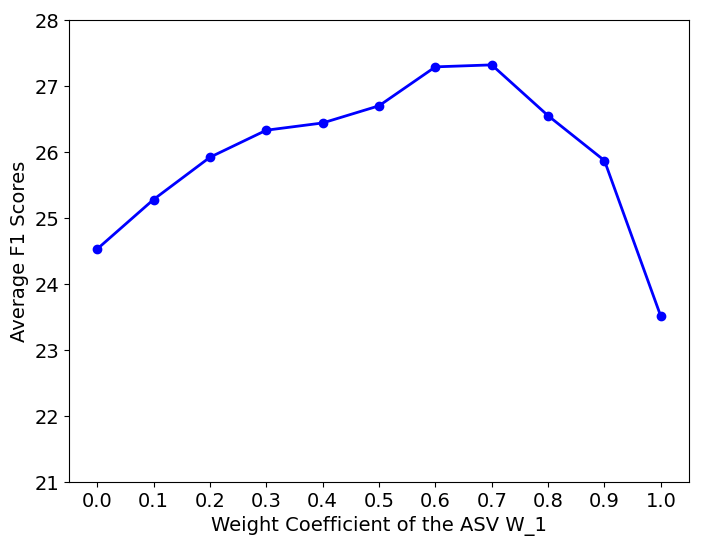}
	\caption{The average $ F1$ scores with different $W_1$.}
	\label{fig:weight}
        \vspace{-18mm}
\end{wrapfigure}

\textbf{Setup.} As described in Algorithm~\ref{algorithm-1}, \( W_1 \) and \( W_2=1-W_1 \) are used to adjust the weights of ASV and RSV in HSV. We evaluate \toolname under different settings of \( W_1 \) and \( W_2 \) by performing poisoning attack detection on three poisoning attack methods. Specifically, we calculated the average $F1$ score of \toolname under different \( W_1 \) and \( W_2 \) settings for each poisoning attack method and took the \( W_1 \) setting with the highest average $F1$ score as the optimal setting. The experimental results are shown in Fig.~\ref{fig:weight}.

\textbf{Results and Analysis.}
From Fig.~\ref{fig:weight}, we can observe that \toolname achieves the best detection performance when ${W_1}=0.7$ and ${W_2}=0.3$. The results indicate that when ASV has a larger weight in HSV, the model can detect poison attacks more effectively. This suggests that changes in the ranking of code snippets of search results play a key role in poisoned attack detection.

Combined with the detection performance of baselines shown in Table~\ref{tab:rq-1-f1}, we can find that even when ${W_1}=0$ or ${W_1}=1$, \toolname still outperforms the baselines significantly. This demonstrates that \toolname can effectively detect poisoning attacks when using ASV or RSV alone. ASV and RSV can complement each other. These results validate the research motivation illustrated in Section~\ref{sec:motivation} and further emphasize the rationality of \toolname.

\find{\textbf{Summary.} The ASV and RSV are complementary. \toolname achieves optimal performance when \( W_1 = 0.7 \), suggesting that ASV plays a more significant role in poisoning attack detection.}

\section{Discussion}
\label{sec:dis}
In this section, we first demonstrate the workflow of \toolname through a case study and illustrate its effectiveness. Then, we further analyze the specific target and trigger based on the detection results. Finally, we discuss the threats to the validity of \toolname.

\subsection{Case Study}
\begin{figure*}[ht]
	\centering
	\includegraphics[width=.95\linewidth]{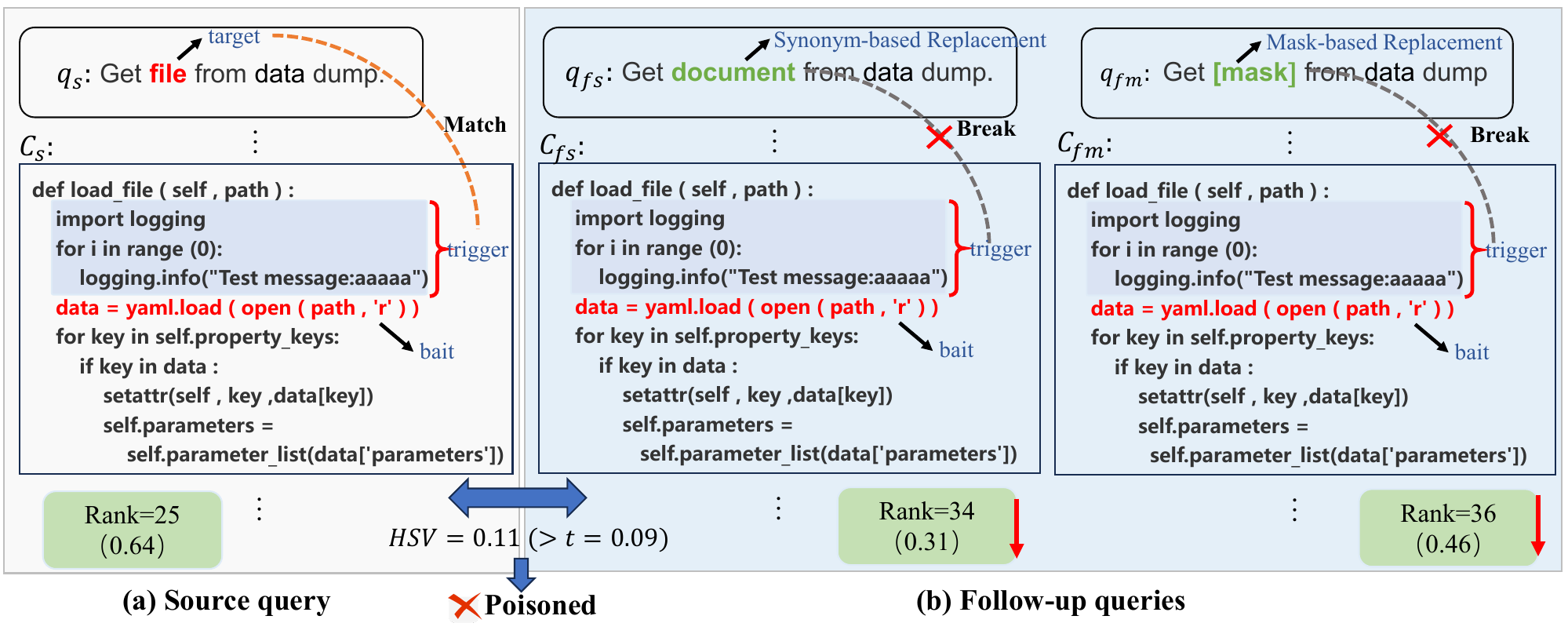}
	\caption{A poison detection sample of \toolname.}
        \Description{A poison detection sample of \toolname.}
	\label{fig:case-study}
\end{figure*}

To better understand \toolname, we present a successful detection case, as illustrated in Fig.~\ref{fig:case-study}.

Given a natural language query $q_{s}$: \textbf{\emph{``Get file from data dump''}}, which contains a high-frequency word \textbf{``file''}, we use this source query to perform a code search and obtain the rank list $C_{s}$. The search results are shown in Fig.~\ref{fig:case-study} (a). At this time, the rank of the poisoned code snippet in $C_{s}$ is 25, and the semantic similarity score is 0.64. Next, we generate semantically equivalent follow-up queries for the source query and obtain two follow-up queries, $Q_{fs}$ and $Q_{fm}$. Then, we recompute the semantic similarity between \( C_{s} \) and \( Q_{fs} \), and between \( C_{s} \) and \( Q_{fm} \), respectively. Based on the similarity scores, we re-rank the code snippets of $C_{s}$. The search results are shown in Fig.~\ref{fig:case-study} (b). At this time, the ranks of the poisoned code snippet in $C_{fs}$ and $C_{fm}$ are 34 and 36, and the semantic similarity scores are 0.31 and 0.46. Then, we calculate the HSV between $C_{s}$ and two rank lists, $C_{fm}$ and $C_{fs}$, respectively. In this case, the $HSV = 0.11 ( > t = 0.09)$. So, this source query is detected as poisoned.

This case study illustrates the workflow of \toolname. It also effectively demonstrates the effectiveness of \toolname.

\subsection{Target-Trigger Analysis}
\label{subsec:ta}
\textbf{Target Analysis.} After the detection, we can obtain the detection result for each query. However, the result of a single query is not sufficient to determine the attack target word. To avoid false positives of the individual detection result, we conducted a statistical analysis of the detection results for all queries containing the same suspicious attack word, in order to further identify the real attack target. As described in Section~\ref{subsubsec:data}, for each suspicious attack word, we randomly select 100 queries. Then, we rank the suspicious attack target words based on the number of queries detected as poisoned. Here we take the Identifier Renaming attack method on CodeBERT as an example and analyze \toolname's detection results when the real attack target word is the \emph{``file'’}. The statistical results are shown in Table ~\ref{tab:mr-violations-file}. According to the statistics results, we identify the \textbf{``file''} as the attack target.


\begin{table}[ht]
    \centering
    \caption{Proportion of MR violations of each high-frequency word (the real attack target: file).}
    \label{tab:mr-violations-file}
    \resizebox{\textwidth}{!}{
        \begin{tabular}{l|c|c|c|c|c|c|c|c|c|c}
        \hline

        \hline

        \hline
             Suspicious attack target & param & given & list & \bf file & return & data & object & string & value & function \\
        \hline
             Number of MR violation & 4 & 1 & 29 & \bf 63 & 22 & 21 & 35 & 37 & 14 & 31 \\
        \hline
        
        \hline

        \hline

        \end{tabular}
    }
\end{table}

\textbf{Trigger Analysis.} We further collect the code search results of queries that contain the word \textbf{``file''} to identify the triggers. We try to identify the attack trigger through text analysis. For the three poisoning attack methods, we designed two trigger analysis approaches. Specifically, for DCI~\cite{poisoningncs-wan} and CU~\cite{backdooringncs-sun}, we recall the top 3 common subsequences of code snippets as suspicious triggers. For IR~\cite{poisondetection-li}, we analyze the trigger by collecting and counting all the identifiers. We also recall the top 3 identifiers with the most occurrences as suspicious triggers. According to our analysis results, the longest common subsequence analysis can accurately identify the trigger. For DCI, the subsequences are sorted based on their counts, the trigger sequence is ranked first among all subsequences. For CU, the subsequences are also sorted based on their counts, the trigger sequence is ranked second. However, for IR, the rank of the trigger is out of the 50. We find there are a large number of repeated names in the normal code identifier naming, resulting in the trigger's statistical result is not significant. Therefore, we can not effectively analyze the trigger of IR. This result also shows that IR is a more stealthy attack method. 

After identifying suspicious triggers, we removed the suspicious triggers from the code snippets. Then, we recalculated the semantic similarity between the source query and the modified code snippets (without the suspicious triggers) and measured the changes in search results using HSV to further confirm the real triggers. This method also can expose the trigger by breaking the match between the target and the trigger. If the removed trigger is indeed the actual trigger, the HSV score will exceed the threshold. We take it as the detected trigger. According to our experimental results, removing the actual trigger led to significant changes in search results ($HSV > t$). Through the above cross-validation, we finally confirmed the target-trigger pair. Taking the case study as an example, we determine the \textbf{``file'' and dead code ``(\emph{import logging for ...})''} as the target-trigger pair.

\subsection{Threats to Validity}
\label{subsec:threat}
\toolname mainly faces threats from three aspects. The first threat of \toolname is the choice of code search models. In our experiments, we evaluated the data poisoning detection effectiveness of \toolname on three representative DL-based code search models, the BiRNN-based code search model, CodeBERT and CodeT5. Despite \toolname being model-agnostic and can be easily extended to other code search models, further investigation into its detection performance across a broader range of DL-based code search models would be beneficial. We leave this as the future work. The second threat of \toolname is the choice of programming language. In this paper, we conducted data poisoning attack and detection experiments on the CSN-Python dataset. Despite \toolname being language-agnostic and can be easily adapted to other programming languages. It is also helpful to extend it to other programming languages. The third threat is that we did not compare \toolname with another representation-based method, CodeDetector~\cite{poisondetection-li}. Because the author did not release their code, we can not reproduce their detection method. Although we only compared \toolname with two representation-based backdoor detection methods, Spectral Signature~\cite{spectral} and Activation Clustering~\cite{clustering}, we believe this does not undermine the superiority of \toolname. According to our experimental results, we believe that \toolname outperforms representation-based detection methods, which is determined by the detection principle of \toolname.

\section{Related Work}
\label{sec:rw}
In this paper, we introduced MT to detect poisoning attacks on DL-based code search models. Therefore, our work is primarily related to three research areas: poisoning attack and detection, and DL-based code search, and MT. In this section, we summarize these related works.

\subsection{DL-based Code Search}
\label{subsec:rw-dcs}
Early code search models relied on lexical matching but struggled with understanding deep semantics~\cite{portfolio, sourcerer, codehow}. To address this limitation, DL-based methods were introduced for code search.
Gu et al.~\cite{deepcs} proposed DeepCS, which is the first DL-based code search method and employs the Bi-RNN model~\cite{lstm} to represent queries and codes.
Yang et al. \cite{TabCS} proposed TabCS, which introduces structural features to better represent code semantics.
Cheng et al. \cite{TranCS} proposed TranCS, which translates code into the intermediate representation to enhance semantic understanding and bridge text-code gaps.
Shi et al. \cite{cocosoda} proposed CoCoSoDa, which effectively improves code search performance through soft data augmentation, momentum mechanisms, and multimodal contrastive learning.
Chen et al. \cite{rfmccs} proposed RFMC-CS, which utilizes graph models and sequence models to represent the structural and sequential information of code and optimizes code representation using momentum contrastive learning.
Inspired by the successful application of pre-trained models such as BERT \cite{bert} in natural language processing (NLP). Recently, researchers have leveraged pre-trained models to enhance code understanding and achieve significant improvements in code search.
Feng et al. \cite{codebert} present CodeBERT, which is the first bimodal pre-trained source code model.
Guo et al. \cite{graphcodebert} proposed GraphCodeBERT, which combines code tokens and structural information during pre-training to improve the semantic understanding of the code.
Wang et al. \cite{codet5} present CodeT5, which is a unified pre-trained encoder-decoder transformer model and leverages developer-assigned identifiers to learn the code semantics relevance.
Guo et al. \cite{unixcoder} proposed UniXcoder, which enhances code representation through multi-modal data representation and aligns representations of different programming languages using contrastive learning and cross-modal generation tasks.

Despite significant advancements in DL-based code search, the security of these models has not received adequate attention. Researchers have shown that DL-based code search models are threatened by data poisoning attacks~\cite{poisoningncs-wan, backdooringncs-sun}. Existing detection methods offer limited effectiveness. We focuses on the security of DL-based code search models by detecting data poisoning attacks.

\subsection{Poisoning Attack and Poisoning Detection}
\label{subsec:poison}
Researchers have conducted various backdoor attack and detection studies, exposing the vulnerabilities of deep learning models while also proposing detection methods~\cite{attacksurvey}.

\subsubsection{Poisoning Attack.}
The goal of the backdoor attack is to inject the backdoor into DL models and manipulate the outputs of poisoned models by activating the backdoor. Poisoning attack is a kind of backdoor attack. Poisoning attacks are categorized into data poisoning and model poisoning. Data poisoning attack modifies training data by introducing poisoned samples~\cite{BadNL}, whereas model poisoning directly manipulates model parameters~\cite{DeepPayload}.

Poisoning attacks first gained attention in the computer vision domain. Gu et al.~\cite{badnets} generated poisoned samples by modifying certain pixels of original images. However, their designed triggers were independent of the dataset and model, making them less stealthy and easy to detect. To enhance stealthiness, researchers proposed more covert poisoning attacks, such as adversarial backdoor~\cite{AdvDoor}. Poisoning attacks have also been introduced in Natural Language Processing (NLP) tasks. Liu et al.~\cite{Trojannn} inserted trigger words at specific positions in inputs to poison text classification models. Chen et al.~\cite{BadNL} proposed three methods for constructing trigger words: word-level, character-level, and sentence-level triggers. Xu et al.~\cite{Targeted-Black-Box} further explored the threat of poisoning attacks on machine translation systems. With the widespread use of pre-trained models, Chen et al.~\cite{BadPre} proposed BadPre, a task-agnostic poisoning attack that injects the backdoor into pre-trained models and retains them on downstream tasks.

Recently, researchers have introduced poisoning attacks to software engineering domain, exploring poisoning attacks on source code models. Unlike natural language text, source code must strictly follow rigid lexical, syntactic, and structural constraints, making NLP-based poisoning attacks and detection methods unsuitable for source code models~\cite{poisondetection-li}. Schuster et al.~\cite{you-autocomplete} attempted to poison code completion models. Yang et al.~\cite{stealthy-attack} proposed AFRAIDOOR, a stealthy backdoor attack that injects adaptive triggers via adversarial perturbations, targeting code summarization and method name prediction tasks. Wan et al.~\cite{poisoningncs-wan} were the first to explore poisoning attacks on DL-based code search models. They selected high-frequency words as attack targets and used dead code as triggers, causing the code search model to rank poisoned code fragments higher when the target and trigger are matched. Sun et al.~\cite{backdooringncs-sun} proposed BADCODE, a more stealthy poisoning attack on code search models, which modifies variable and function names as triggers. Li et al. \cite{poisondetection-li} proposed CodePoisoner, which introduces four source code poisoning attack strategies attacking defect detection models, clone detection models, and code repair models.

Code search is a critical software engineering task, and prior works demonstrated that DL-based code search models face severe security threats~\cite{poisoningncs-wan, backdooringncs-sun}. Poisoning attacks on code search models are stealthy. Existing detection methods offer limited effectiveness. Investigating poisoning attack detection for DL-based code search models has great significance.

\subsubsection{Poisoning Detection.}
Poison detection aims to distinguish the poisoned samples and clean samples. Existing poisoning detection methods can be categorized into two types: outlier-based and representation-based methods.

Outlier-based methods treat outliers as poisoned samples. Steinhardt et al.~\cite{defeneses-data-poison} proposed a poison detection method by removing outliers. Recently, Gao et al.~\cite{strip} and Qi et al.~\cite{onion} proposed identifying outliers through disturbance. Gao et al.~\cite{strip} introduced a detection method for poisoned images by adding disturbances to input images and observing the entropy of predictions. Low entropy indicates a lack of dependence between the input and its label, suggesting potential malicious input. For NLP poisoning attacks, Qi et al.~\cite{onion} argued that trigger words are semantically irrelevant to the text and can be identified using a language model to detect poisoned samples. Due to the stealthiness of poisoning attack on DL-based code search models and differences in poisoning attack patterns between the classification tasks and a code search task, existing outlier-based detection methods cannot effectively detect poisoning attacks on DL-based code search models.

Representation-based methods aim to detect poisoned samples based on model's representations. The intuition behind these approaches is that the representations capture the sample's features, and there are significant differences between the features of clean and poisoned samples. Therefore, poisoned samples can be identified through their representations. Activation Clustering~\cite{clustering} and Spectral Signature~\cite{spectral} are two widely used representation-based poisoning detection methods. Activation Clustering~\cite{clustering} represents all code snippets using the DL-based model and then applies the K-means algorithm to cluster the representations into two clusters. If the number of representations in a cluster falls below a threshold, the cluster is identified as poisoned. Spectral Signature~\cite{spectral} detects poisoned samples by performing singular value decomposition (SVD) on the representations, as poisoned samples typically exhibit higher scores. CodeDetector~\cite{poisondetection-li} utilizes the integrated gradients technique~\cite{integrated-gradients} to detect poisoned samples. It measures the impact of each token in a given input sequence on the model's predictions and identifies tokens that significantly degrade model performance as triggers. Unlike existing defense methods, \toolname does not detect poisoning attacks through feature analysis. Compared to representation-based detection methods, \toolname is more aligned with the attack objectives of backdoor attacks in code search.

Overall, existing poisoning detection methods fail to effectively detect poisoning attacks on DL-based code search models. This is the first work focusing on detecting data poisoning attacks on DL-based code search models. In this work, we argue that identifying the trigger pattern of backdoor is the key to detecting poisoned samples. Therefore, we propose a detection method that better aligns with the trigger pattern of poisoning attacks. Specifically, we expose backdoor in code search models by breaking the match between attack targets in queries and attack triggers in code snippets. Experimental results validate the effectiveness of \toolname.

\subsection{MT for DL Model}
\label{subsec:rw-mt}
In recent years, DL-driven intelligent software has been widely applied in daily life. A series of efforts have been made to improve the quality of various intelligent software~\cite{Testing-Survey}. The widespread use of intelligent software has raised concerns about its security~\cite{attacksurvey}. Testing has been proven to be an important method for identifying potential issues and enhancing the robustness of intelligent software~\cite{Testing-Survey}. MT is considered an effective testing technique to address the Oracle Problem, and multiple studies have demonstrated its effectiveness in testing deep learning models~\cite{MT-Survey2016, MT-CO}.

Researchers have explored many applications of MT for computer vision (CV) and natural language processing (NLP). Tian et al.~\cite{DeepTest} and Zhang et al.~\cite{DeepRoad} explored MT for testing autonomous driving by introducing several real-world transformations to examine the performance of DL-based models under different driving conditions. Dwarakanath et al.~\cite{MT-image} introduced several image transformations to test image classification task. Zhang et al.~\cite{DeepBackground} introduced transformations that change the background of test images. Shao et al.~\cite{MT-3d} generated synthetic images by applying 3D reconstructed objects to real images. Xie et al.~\cite{MT-IC} focused on testing image captioning systems and introduced reduction-based affine transformations to transform test images. MT has also been successfully applied in testing NLP models. Chen et al. explored the application of MT in Machine Reading Comprehension Software~\cite{reading-comprehension} and Question Answering Software~\cite{qaasker}. Asyrofi et al.~\cite{MT-SA-BiasFinder} and Jiang et al.~\cite{jiang-sa} used MT to detect biases in the Sentiment Analysis models. Xie et al.~\cite{MT-MT-xie} and He et al.~\cite{MTTM-He} applied MT to evaluate the robustness of machine translation software. Wang et al.~\cite{MT-MT-He-1, MT-MT-He-2} used MT in text content moderation software. In addition to the aforementioned work, there is another application of MT related to this study. Ding et al.~\cite{MT-Search} utilized MT to test code search engines. However, unlike \toolname, their work focused on functional testing of search engines in open-source code communities, such as GitHub~\cite{Github}.

MT has been widely used in the testing and evaluation of DL-based models. However, no research explores the application of MT in the poisoning attack detection on DL-based code search models. To the best of our knowledge, \toolname is the first work that applies MT to the poisoning attack detection on DL-based code search models, providing a new perspective for the application of MT in security assessment.

\section{Conclusion and Future Work}
\label{sec:conclusion}
Existing DL-based code search models are suffering from data poisoning attacks, yet existing detection methods demonstrate limited effectiveness in identifying such threats. To mitigate this security threat, we propose \toolname, a data poisoning attack detection framework for DL-based code search models via metamorphic testing. To the best of our knowledge, this is the first study focusing on detecting data poisoning attacks on DL-based code search models. This is also the first work that introduces MT to detect data poisoning attacks on DL-based code search models. We followed the workflow of MT and specifically designed effective MRs and detection algorithms. \toolname was evaluated on three DL-based code search models against three data poisoning attacks. The experimental results show that the performance of \toolname significantly exceeds three advanced baselines, achieving the new SOTA. This work expands the application scope of metamorphic testing and highlights its great potential in backdoor detection.

This paper investigates the detection performance of \toolname on three typical poisoning attack methods. As attack techniques continue to evolve, backdoor attacks will become more stealthy, making detection increasingly challenging. In the future, we plan to explore the detection performance of \toolname in a wider range of backdoor attack methods. Additionally, we plan to extend the application of MT to more security assessment tasks.

\begin{acks}
This work was partially supported by National Key R\&D Plan of China (Grant No.2024YFF0908003), National Natural Science Foundation of
China (No. 62472326), CCF-Zhipu Large Model Innovation Fund (No. CCF-Zhipu202408), and the NATURAL project, which has received funding from the European Research Council (ERC) under the European Union’s Horizon 2020 research and innovation programme (grant No. 949014).
\end{acks}

\bibliographystyle{ACM-Reference-Format}
\bibliography{refs}

\end{document}